\documentclass[prl,nofootinbib,floats,aps,superscriptaddress,twocolumn]{revtex4}

\usepackage[hypertex]{hyperref}
\usepackage{graphicx} 
\usepackage{xspace,amsmath}

\newcommand{\spsi}{S_{\psi K}}
\newcommand{\spsph}{S_{\psi \phi}}

\newcommand{\asls}{a^s_{\rm SL}}
\newcommand{\afss}{a^s_{\rm fs}}
\newcommand{\asld}{a^d_{\rm SL}}

\newcommand{\aslb}{a^b_{\rm SL}}

\def\Eq#1{Eq.~(\ref{#1})}

\newcommand{\beq}{\begin{equation}}
\newcommand{\eeq}{\end{equation}}
\newcommand{\beqa}{\begin{eqnarray}}
\newcommand{\eeqa}{\end{eqnarray}}
\newcommand{\nn}{\nonumber}
\begin{document}

\pagestyle{plain}  

\title{\boldmath Implications of the dimuon CP asymmetry in $B_{d,s}$ decays}

\author{Zoltan Ligeti}
\affiliation{Ernest Orlando Lawrence Berkeley National Laboratory,
University of California, Berkeley, CA 94720}

\author{Michele Papucci}
\affiliation{Institute for Advanced Study, Princeton, NJ 08540}

\author{Gilad Perez}
\affiliation{\mbox{Department of Particle Physics and Astrophysics,
Weizmann Institute of Science, Rehovot 76100, Israel}}

\author{Jure Zupan}
\affiliation{Faculty of Mathematics and Physics, 
University of Ljubljana Jadranska 19, 1000 Ljubljana, Slovenia}
\affiliation{Josef Stefan Institute, Jamova 39, 1000 Ljubljana, Slovenia}
\affiliation{SISSA, Via Bonomea 265, I 34136 Trieste, Italy}


\begin{abstract}

The D\O\ Collaboration reported a $3.2\,\sigma$ deviation from the standard
model prediction in the like-sign dimuon asymmetry. Assuming that new physics
contributes only to $B_{d,s}$ mixing, we show that the data can be analyzed
without using the theoretical calculation of $\Delta\Gamma_s$, allowing for
robust interpretations. We find that this framework gives a good fit to all
measurements, including the recent CDF $S_{\psi\phi}$ result. The data allow
universal new physics with similar contributions relative to the SM in the $B_d$
and $B_s$ systems, but favors a larger deviation in $B_s$ than in $B_d$ mixing.
The general minimal flavor violation framework with flavor diagonal CP violating
phases can account for the former and remarkably even for the latter case. This
observation makes it simpler to speculate about which extensions with general
flavor structure may also fit the data.

\end{abstract}

\maketitle



In the last decade an immense amount of measurements determined that the
standard model (SM) is responsible for the dominant part of flavor and CP
violation in meson decays.  However, in some processes, mainly related to $B_s$
decays, possible new physics (NP) contributions are still poorly constrained,
and motivated NP scenarios predict sizable deviations from the SM.
Recently the D\O\ Collaboration reported a measurement of the like-sign dimuon
charge asymmetry in semileptonic $b$ decay with improved precision~\cite{D0},
\begin{equation}\label{new}
\aslb \equiv \frac{N_b^{++} - N_b^{--}}{N_b^{++} + N_b^{--}}
  = -(9.57 \pm 2.51 \pm 1.46) \times 10^{-3},
\end{equation}
where $N_b^{++}$ is the number of $b\bar b\to \mu^+\mu^+X$ events (and similarly
for $N_b^{--}$).  This result is $3.2\sigma$ from the quoted SM prediction,
$\left(\aslb\right)^{\rm SM} = (-2.3^{+0.5}_{-0.6}) \times
10^{-4}$~\cite{Lenz:2006hd}. At the Tevatron both $B^0_d$ and $B^0_s$ are
produced, and hence $\aslb$ is a linear combination of the two
asymmetries~\cite{D0}
\beq
\aslb = \left(0.506\pm0.043\right)\asld 
  + \left(0.494\pm0.043\right)\asls\,.\label{weight}
\eeq

The above result should be interpreted in conjunction with three other
measurements: (i) the $B_d$ semileptonic asymmetry, measured by the $B$
factories, $\asld=-(4.7\pm 4.6) \times 10^{-3}$~\cite{Barberio:2008fa}; (ii) the
flavor specific asymmetry measured from time dependence of $B^0_s\to \mu^+ D_s^-
X$ decay and its CP conjugate, $\afss = -(1.7\pm 9.1\pm1.5)\times
10^{-3}$~\cite{flavorspecific}; and (iii) the measurements of $\Delta \Gamma_s$
and $\spsph$ (the CP asymmetry in the CP-even part of the $\psi\phi$ final state
in $B_s$ decay)~\cite{D0betas,CDFbetas,Tevbetas,CDFbetasnew}.  Here $\Delta
\Gamma_s = \Gamma_L - \Gamma_H$, is the width difference of the heavy and light
$B_s$ mass eigenstates. If CP violation is negligible in the relevant tree-level
decays, then $\afss = \asls$.  The SM predictions for the asymmetries $\asld$
and $\asls$ are negligibly small, beyond the reach of the Tevatron
experiments~\cite{LLNP,Beneke:2003az,Ciuchini:2003ww}. If the evidence for the 
sizable dimuon charge asymmetry in Eq.~(\ref{new}) is confirmed, it would
unequivocally point to CP violation beyond the CKM mechanism of the SM. 

The present experimental uncertainties of $\asld$ and $\asls$ separately are
larger than that of  their combination, $\aslb$.  Thus, from Eq.~(\ref{new})
alone it is not clear if the tension with the SM is in the $B_d$ or in the $B_s$
system.  Bounds from other observables imply (see below) that new physics
contributions in $B_d$ mixing with a generic weak phase cannot exceed roughly
20\% of the SM, while in $B_s$ mixing much larger NP contributions are still
allowed.

We focus on interpreting the data \emph{assuming} that the above measurements
are associated with new CP violating physics which contributes to $B_{d,s}$
mixing, while its contribution to CP violation in tree-level decay amplitudes is
negligible. Under this assumption the D\O\ result in Eq.~(\ref{new}) is
correlated with the Tevatron measurements of $\spsph$~\cite{Ligeti:2006pm} (and
$\Delta \Gamma_s$). These measurements provide nontrivial tests of our
hypothesis (see~\cite{Kagan:2009gb} for relaxing these assumptions).  Neglecting the 
small SM contribution to $\spsph$, the following relation
holds between experimentally measurable quantities~\cite{Grossman:2009mn}
\beq\label{aslsppcor}
\asls = - \frac{|\Delta \Gamma_s|}{\Delta m_s}\, 
  {\spsph}\,\big/ {\sqrt{1-\spsph^2}}\,,
\eeq
where $\Delta m_s \equiv m_H - m_L$.  Using the new measurement in \Eq{new}
together with \Eq{weight}, the above relation implies
\begin{equation}
\left|\Delta \Gamma_s\right| \simeq -\Delta m_s 
  \big(2.0\,\aslb-1.0\,\asld\big)\,
  \sqrt{1-\spsph^2}\, \big/\, \spsph \,.
\label{corr}
\end{equation}
For simplicity we do not display the ${\cal O}\left(10\%\right)$ uncertainties
of the two numerical factors.  The CDF and D\O\ time-dependent $B_s\to \psi\phi$
analyses provide a measurement of $\Delta \Gamma_s$ vs.\ $\spsph$.  Hence all
quantities in Eq.~(\ref{corr}) are constrained, and our analysis can be
performed without the theoretical prediction of $\Delta
\Gamma_s$~\cite{Grinstein:2001nu}, using its determination from data instead.

Using the measured values of $\Delta m_s$ and $a_{\rm SL}^{b,d}$, we find
\beq 
|\Delta \Gamma_s| \sim \big[(0.28 \pm 0.15)\,{\rm ps}^{-1} \big]\,
  \sqrt{1-\spsph^2}\, \big/ \spsph\,.\label{correxp}
\eeq
The recent CDF~\cite{CDFbetasnew} and D\O~\cite{D0betas} results give best
fit values around $(\Delta \Gamma_s,\, \spsph) \sim (\pm 0.15\,{\rm ps}^{-1},\,
0.5)$.  This shows that the new $\aslb$ measurement in Eq.~(\ref{new}) is
consistent with the data on $\Delta \Gamma_s$ and $\spsph$.  This consistency is
a nontrivial test of the assumption that NP contributes only to neutral meson
mixing.

New physics in the mixing amplitudes of the $B_{d,s}$ mesons can in general be
described by four real parameters, two for each neutral meson system,
\beq\label{hsigma}
M_{12}^{d,s} = \big(M_{12}^{d,s}\big)^{\rm SM}
  \left(1 + h_{d,s}\, e^{2i\sigma_{d,s}}\right) .
\eeq
We denote  by $M_{12}^{q}$ ($\Gamma_{12}^q$) the dispersive (absorptive) part of
the $B^0_q-\bar B^0_q$ mixing amplitude and SM superscripts denote the SM values
(for quantities not explicitly defined here, see Ref.~\cite{Anikeev:2001rk}). 
This modifies the SM predictions for some observables used to constrain $h_q$
and $\sigma_q$ as
\beqa\label{par}
\Delta m_q &=& \Delta m_q^{\rm SM}\,
  \big|1+h_q e^{2i\sigma_q}\big| \,, \nn\\
\Delta\Gamma_s &=& \Delta\Gamma_s^{\rm SM}\,
  \cos \big[\arg\big(1+h_s e^{2i\sigma_s}\big)\big] \,,\nn\\
A^q_{\rm SL} &=& {\rm Im}\, \big\{ \Gamma_{12}^q / \big[ M_{12}^{q,{\rm SM}}
  (1+h_q e^{2i\sigma_q}) \big] \big\} \,,\nn\\
\spsi &=& \sin \big[2\beta + \arg\big(1+h_d e^{2i\sigma_d}\big)\big]\,, \nn\\
\spsph &=& \sin \big[ 2\beta_s - \arg \big(1+h_s
    e^{2i\sigma_s}\big)\big]\,.
\eeqa
Here $\beta_s 
= \arg[-(V_{ts}V_{tb}^*)/(V_{cs}V_{cb}^*)]
= (1.04 \pm 0.05)^\circ$ is an angle of a squashed unitarity triangle.

As already discussed, the new D\O\ measurement directly correlates the possible
NP contributions in the $B_d$ and $B_s$ systems [see \Eq{weight}].  In order to
quantitatively assess our NP hypothesis we perform a global fit using the {\tt
CKMfitter} package~\cite{ckmfitter} to determine simultaneously the NP
parameters $h_{d,s}$ and $\sigma_{d,s}$, as well as the $\bar\rho$ and
$\bar\eta$ parameters of the CKM matrix. 

The results presented here use the post-Beauty2009 {\tt CKMfitter} input
values~\cite{ckmfitter}, except for the lattice input parameters where we
use~\cite{Laiho:2009eu}, and the most recent experimental data.  For $\spsph$ 
vs.\ $\Delta \Gamma_s$, we use the 2.8\,fb$^{-1}$ 2d likelihood of
D\O~\cite{D0betas} and the 5.2\,fb$^{-1}$ 1d likelihood of the recent CDF
measurement~\cite{CDFbetasnew} (the 2d likelihood is not available); these fits
are done without assumptions on the strong phases.  As already mentioned,
neither the CDF nor the D\O\ result gives a significant tension in the fit, so
we expect that a real 2d Tevatron combination of the ICHEP~2010 results
\cite{CDFbetasnew,D0betasnew} will not alter our results significantly.  For the
results presented here, we marginalize over $|\Gamma_{12}^s|$ in the range
$0-0.3\,{\rm ps}^{-1}$, finding that the data prefer values for
$\Delta\Gamma_s$ about 2.5 times larger than the prediction~\cite{Lenz:2006hd}.
If we use the theory prediction, our conclusions about NP do not change
substantially, but the goodness of fit is reduced significantly.

\begin{figure}[t]
\includegraphics[width=0.95\columnwidth]{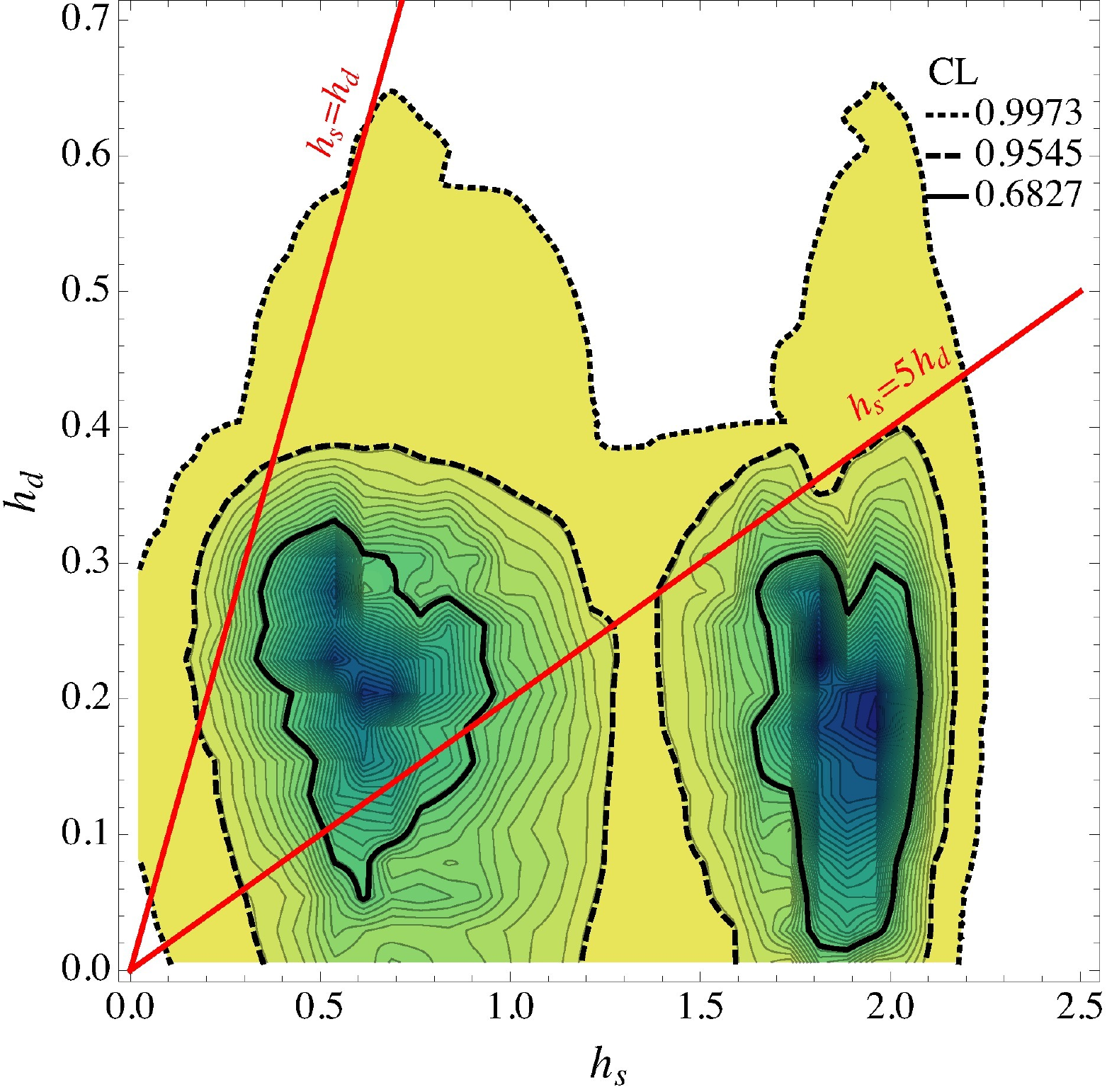}
\caption{The allowed range of $h_s$ and $h_d$ from the combined fit.  The solid,
dashed, and dotted contours show 1$\sigma$, 2$\sigma$, and $3\sigma$, respectively.}
\label{fig_hshd}
\end{figure}

\begin{figure*}[t]
\includegraphics[width=0.95\columnwidth]{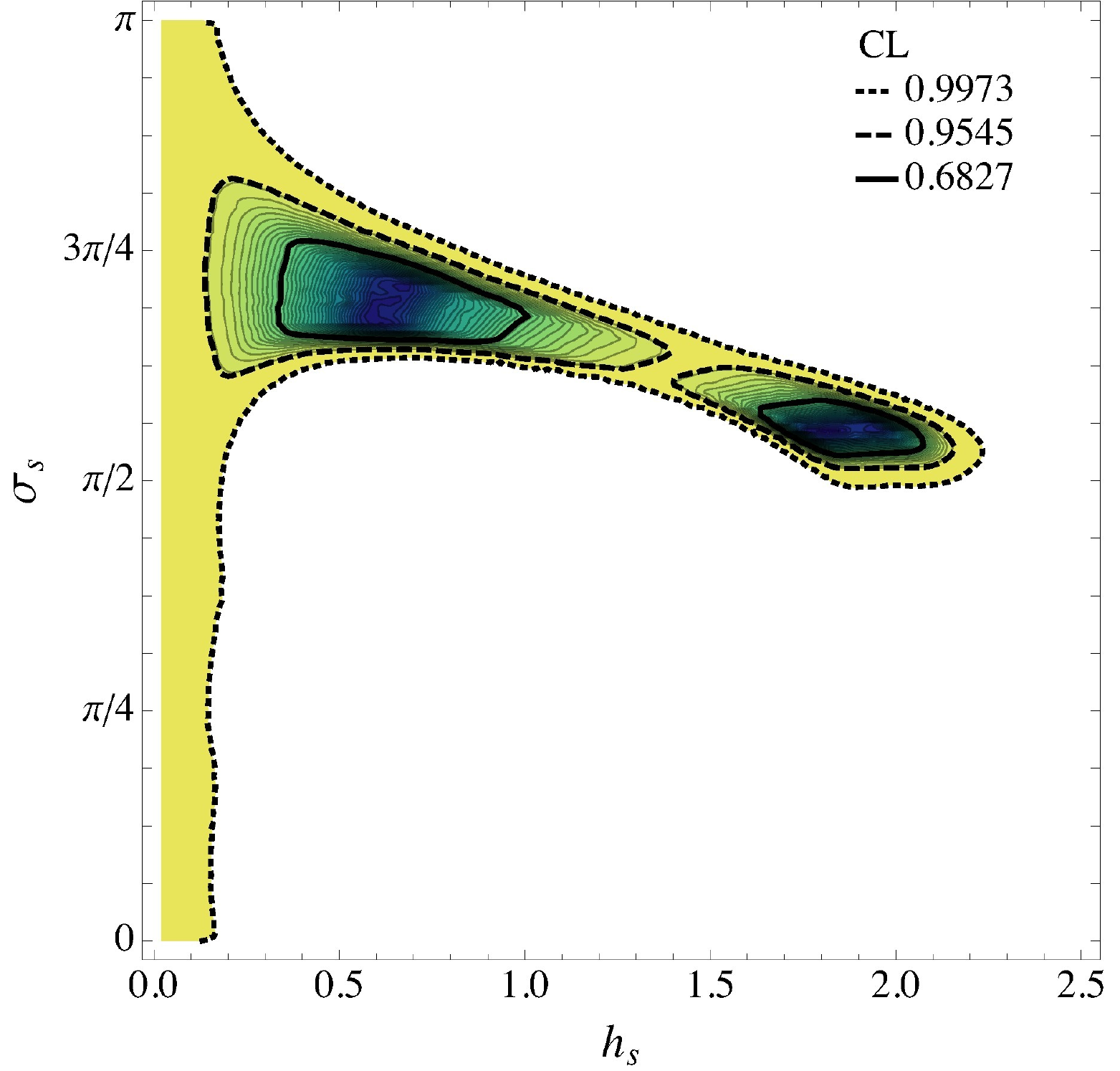} \hfil
\includegraphics[width=0.95\columnwidth]{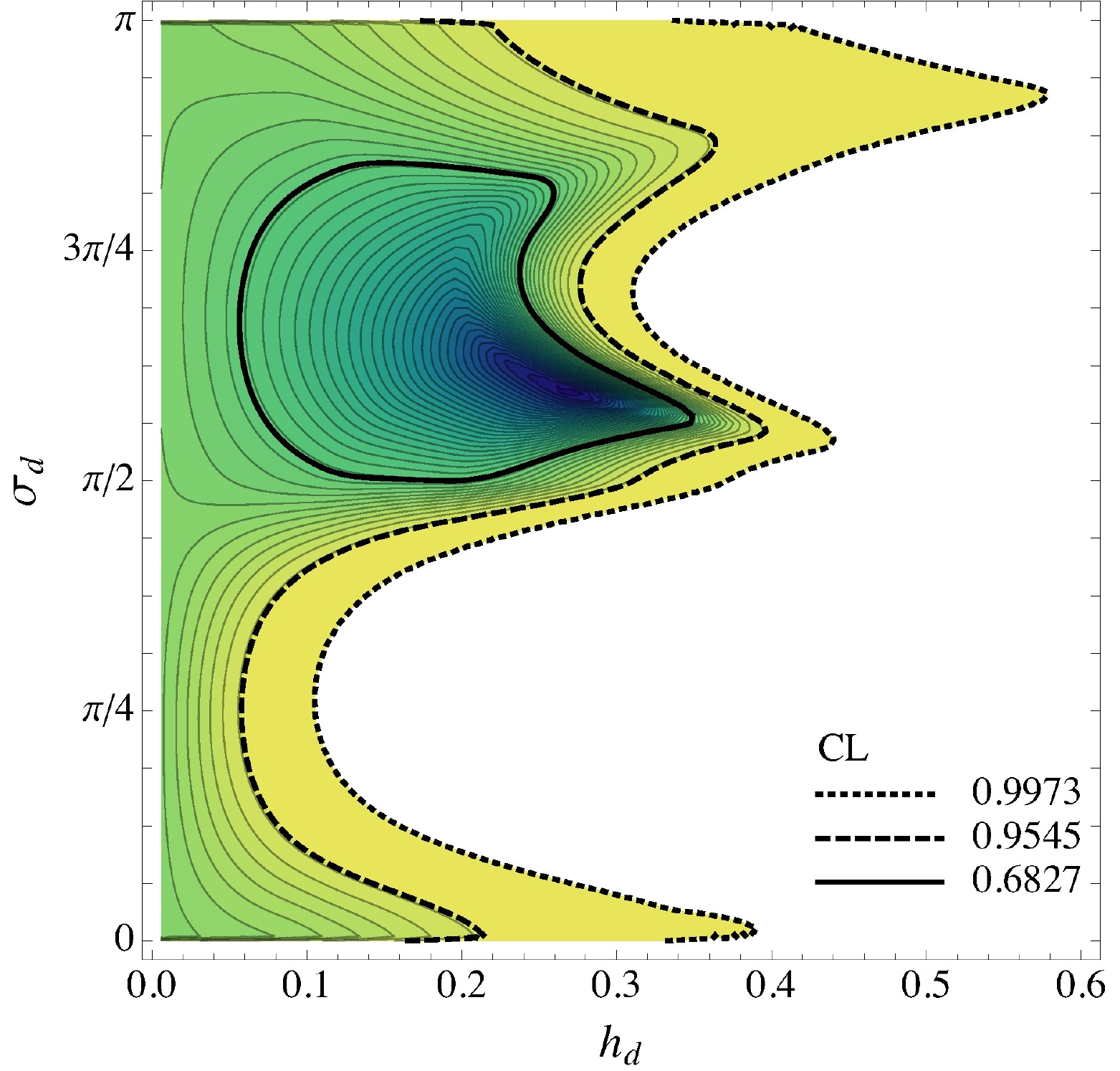}
\caption{The allowed ranges of $h_s,\sigma_s$ (left) and $h_d,\sigma_d$ (right)
from the combined fit to all four NP parameters.}
\label{fig_hsig}
\end{figure*}

Figure~\ref{fig_hshd} shows the results of the global fit projected onto the
$h_d-h_s$ plane with 1$\sigma$ (solid), 2$\sigma$ (dashed), and 3$\sigma$
(dotted) contours.  We find that the data show evidence for disagreement with
the SM or, differently stated, the no NP hypothesis $h_s=h_d=0$ is disfavored
at the 3.3$\sigma$ level.  Figure~\ref{fig_hsig} shows the $h_s-\sigma_s$ and
$h_d-\sigma_d$ fits.  The two best fit regions are for $h_s\sim 0.5$ and $h_s\sim 1.8$ with sizable NP
phases, $\sigma_s \sim 120^\circ$ and $\sigma_s \sim 100^\circ$ respectively.  Here the point $h_s=0$ is disfavored at only
$2.6\sigma$, since $h_s$ and $h_d$ are correlated.  In the $h_d-\sigma_d$ case
the data is consistent with no new physics contributions in $B_d-\bar B_d$
mixing ($h_d=0$) below the 2$\sigma$ level.

To interpret the pattern of the current experimental data in terms of NP models,
one should investigate if NP models that respect the SM approximate $SU(2)_q$
symmetry are favored (in the SM this is due to the smallness of the masses in
the first two generations and the  smallness of the mixing with the third
generation quarks), or if a hierarchy, such as $h_s \gg h_d$, is required. In
Fig.~\ref{fig_hshd} we show the $h_d=h_s$ line, which makes it evident that
while $h_d=h_s$ is not disfavored, most of the favored parameter space has $h_s
> h_d$. Actually, a non-negligible fraction of the allowed parameter space
corresponds to $h_s \gg h_d$, as indicated by the $h_s=5 h_d$ line on
Fig.~\ref{fig_hshd}. 

\begin{figure}[b]
\includegraphics[width=0.95\columnwidth]{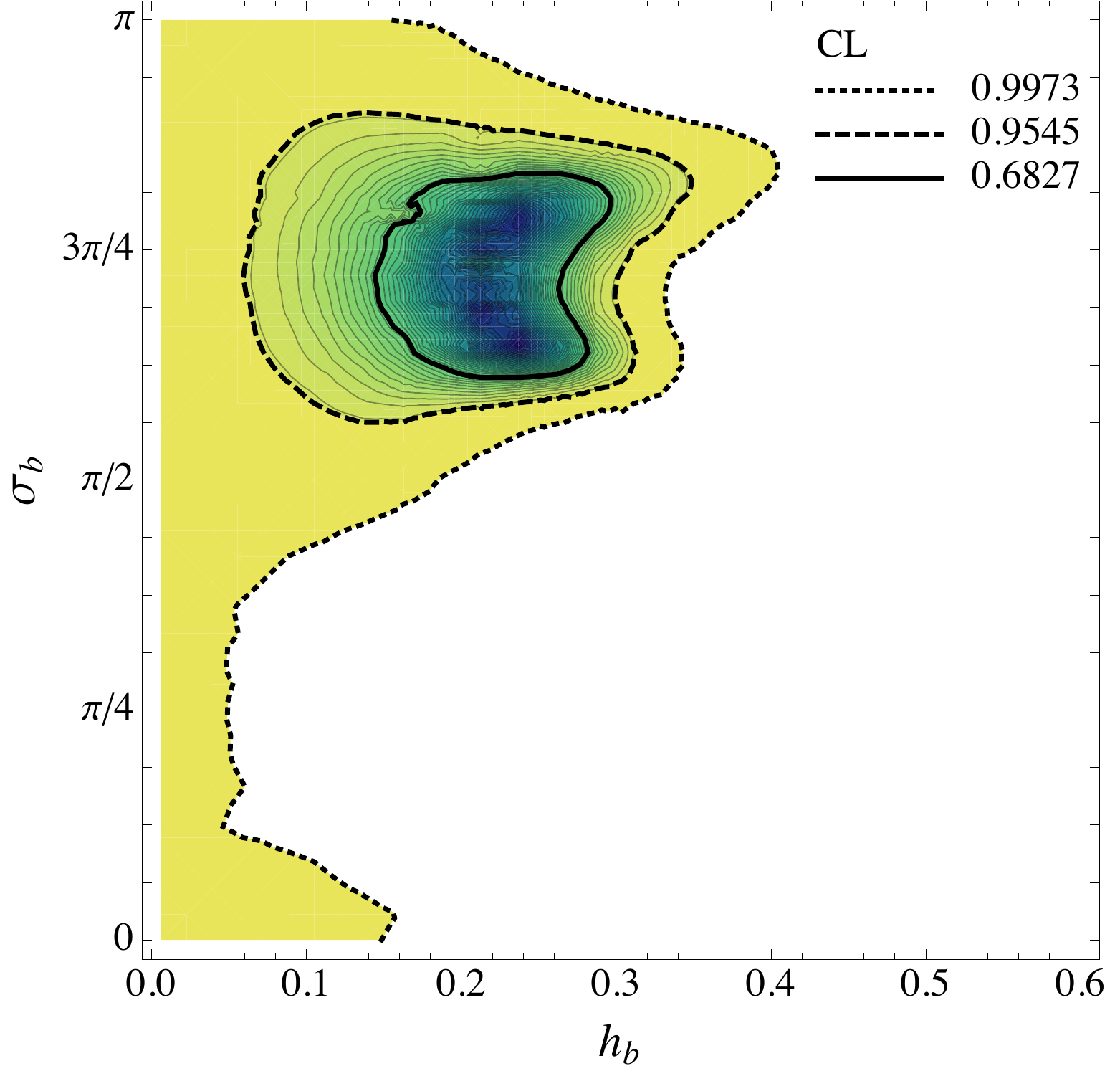}
\caption{The allowed $h_b,\sigma_b$ range assuming $SU(2)$ universality.}
\label{fig_hbsigb}
\end{figure}

A particularly interesting NP scenario is to assume $SU(2)_q$ universality
($q=s,d$), defined as
\beq 
h_b \equiv h_d = h_s\,, \qquad 
  \sigma_b \equiv \sigma_d = \sigma_s\,.\label{SU2}
\eeq 
The relevant $h_b-\sigma_b$ plane is shown in Fig.~\ref{fig_hbsigb}. The best
fit region, near $h_b \sim 0.25$ and $\sigma_b \sim 120^\circ$, is obtained as a
compromise between the Babar and Belle bounds in the $B_d$ system and the
tensions in the Tevatron $B_s$ data with the SM predictions. This compromise
mostly arises from the different magnitudes of $h_{d,s}$: while the best fit
$h_d$ value is a few times smaller than the best fit $h_s$ value, the best fit
values of the phases $\sigma_{d,s}$ are remarkably close to each other, as can
be seen in Fig.~\ref{fig_hsig}.  Note that while the SM limit, $h_b=0$, is
obtained at less than $3\sigma$ CL, the goodness of the fit is significantly
degraded compared with the non-universal case.

We now move to interpreting the above results, assuming that the dimuon
asymmetry is indeed providing evidence for deviation from the SM. Interestingly,
without restricting our discussion to a specific model, we can still make the
following general statements:

(i) The present data support the hypothesis that new sources of CP violation are
present and that they contribute mainly to $\Delta F=2$ processes via the mixing
amplitude. As is well known, these processes are highly suppressed in the SM.

(ii) The SM extensions with $SU(2)_q$ universality, where the new contributions
to $B_d$ and $B_s$ transition are similar in size (relative to the SM),  can
accommodate the data but are not  the most preferred scenarios experimentally.
Universality is expected in a large class of well motivated models with
approximate $SU(2)_q$ invariance, for instance when flavor transitions are
mediated by the third generation sector~\cite{NMFV}. The case where the NP
contributions are $SU(2)_q$ universal (see \Eq{SU2} and Fig.~\ref{fig_hbsigb}) 
is also quite generically obtained in the minimal flavor violation (MFV)
framework~\cite{MFV} where new diagonal CP violating phases are
present~\cite{Colangelo:2008qp, Kagan:2009bn}. In an effective theory approach
such a contribution may arise from the four-quark operators
$O_1^{b q} = \bar b^\alpha_{L} \gamma_\mu q^\alpha_{L}\,
\bar b^\beta_{L} \gamma_\mu q^\beta_{L}$, 
$O_2^{b q} = \bar b^\alpha_{R} q^\alpha_{L}\, \bar b^\beta_{R} q^\beta_{L}$, 
$O_3^{b q} = \bar b^\alpha_{R} q^\beta_{L}\, \bar b^\beta_{R} q^\alpha_{L}$,
suppressed by scales $\Lambda_{{\rm MFV};1,2,3}$, respectively.
We find that the data require
\beq
\Lambda_{\rm MFV;1,2,3} \gtrsim \{8.8,\ 13\, y_b,\ 6.8\,y_b\}\,
  \sqrt{0.2/h_b}\ {\rm TeV}\,.
\eeq
If the central value of the measurement in Eq.~(\ref{new}) is confirmed, this
inequality would become an equality. Note that the dependence on the bottom
Yukawa, $y_b$, is not shown for $\Lambda_{{\rm MFV};1}$, since sizable CP
violation in this case requires resummation of large effective bottom Yukawa
coupling~\cite{Kagan:2009bn,TASI10}. In general the presence of flavor diagonal
phases could contribute to the neutron electric dipole
moment~\cite{Mercolli:2009ns}. However, this effect arises from a different
class of operators and requires a separate investigation. Another interesting
aspect of these flavor diagonal phases is that there are examples where these
can contribute to the generation of matter-antimatter asymmetry, another issue
which deserves further investigation.

(iii) While case (ii) is not excluded by the data, Fig.~\ref{fig_hshd} shows
that most of the allowed parameter space prefers $h_s> h_d$.  This raises the
following question: \emph{What kind of new physics can generate a large breaking
of the approximate $SU(2)_q$ symmetry without being excluded by CP violation in
the $K$ or $D$ systems?}  Remarkably, even this case can be accounted for by the
general MFV (GMFV) framework~\cite{Kagan:2009bn}.  Consider models where
operators with $O_4$-type chiral and color structure (defined in
\cite{Bona:2007vi}) are the dominant ones.  This may be possible because their
contributions are RGE enhanced. An example of such an operator is (similar
$O_5$-type operators are typically suppressed compared to the $O_4$-type
ones)
\beq
O^{\rm NL}_4 = {c\over \Lambda^2_{{\rm MFV};4}}
  \big[\bar Q_3 (A_d^m A_u^n Y_d)_{3i} d_i\big] 
  \big[\bar d_3 (Y_d^\dagger A_d^{l,\dagger} A_u^{p,\dagger})_{3i} Q_i\big] .
\label{O4G}
\eeq
Here $A_{u,d}\equiv Y_{u,d} Y_{u,d}^\dagger$ and $n,m,l,p$ are integer powers
and $c$ is an ${\cal O}(1)$ complex number.  We focus on the nonlinear MFV
regime, where the contributions of higher powers of the Yukawa couplings are
equally important, so a resummation of the third generation eigenvalues is
required (both for the up and down Yukawas), due to large logarithms or large
anomalous dimensions.  In Eq.~(\ref{O4G}) we adopt a linear formulation where
the resummation of the third generation is not manifest;
see~\cite{Kagan:2009bn,TASI10} for a more rigorous treatment.  Such operators
can carry a new CP violating phase and may contribute dominantly to $b\to s$ and
\emph{not} to $b\to d$ transition, because of the chiral suppression induced by
$Y_d$.  We find that the data requires%
\beq
\Lambda_{{\rm MFV};4} \gtrsim 13\, y_b\, \sqrt{\frac{m_s}{m_b}\, 
  \frac{0.5}{h_s}}\ {\rm TeV}
  \approx 2\,y_b\, \sqrt{\frac{0.5}{h_s}}\ {\rm TeV}\,.
\eeq
Thus, remarkably, $h_s\gg h_d$ can arise in MFV models with flavor diagonal CP
violating phases, where large chirality flipping sources exist at the TeV
scale.  Such models have not been studied in great detail, but possible
interesting examples are supersymmetric extensions of the SM at large $\tan
\beta$~\cite{Bobeth:2002ch} or warped extra dimension models with MFV structure
in the bulk~\cite{Shining}. We finally note that the operator $O^{\rm NL}_4$
predicts contributions to the $B_d$ system suppressed by $m_d/m_s\sim 5\%$,
which may be accessible in the near future and provide a direct test for the
above scenario.

(iv) The fact that the data can be accounted for within the MFV framework makes
it clear that it can be accommodated in models with even more general flavor
structure~\cite{Randall:1998te, Dobrescu:2010rh}.  Several conditions need to be
met, though. For instance, the operators $O_{2,3,4}$ require large chirality
violating sources in addition to the CP violating phases, which are generically
strongly constrained by  neutron electric dipole moment and $b\to s\gamma$. 
Contributions to the $O_1$ operator from $SU(2)_{\rm w}$ invariant new physics,
on the other hand, are constrained by CP violation in $D-\bar D$ mixing. They
may also induce observable $\Delta t=1$ and $\Delta t=2$ top flavor violation at
the LHC~\cite{Fox:2007in,Gedalia:2010mf}.


{\bf Acknowledgments:} 
During the completion of this work, Ref.~\cite{Buras:2010mh} appeared with
partial overlaps with our ideas.
We thank Ben Grinstein, Yonit Hochberg, Jernej Kamenik, Heiko Lacker and Yossi
Nir for useful discussions.
ZL thanks the Aspen Center for Physics for hospitality while this work was
completed. The work of ZL was supported in part by the U.S.\ Department of
Energy under contract DE-AC02-05CH11231.
The work of MP was supported by the NSF under the grant PHY 0907744.
GP~is the Shlomo and Michla Tomarin career development chair, and is supported
by the Israel Science Foundation (grant \#1087/09), EU-FP7 Marie Curie, IRG
fellowship and the Peter \& Patricia Gruber Award.
JZ is supported by the EU Marie Curie IEF Grant.\ no.\ PIEF-GA-2009-252847 and
by the Slovenian Research Agency.


\end{document}